\documentclass[aps,prl,twocolumn,showpacs]{revtex4}
\usepackage{graphicx}

\begin{document}

\title{Bulk Evidence for s-Wave Pairing Symmetry in the n-Type Infinite-Layer Cuprate $Sr_{0.9}La_{0.1}CuO_{2}$}

\author{Z. Y. Liu$^1$, H. H. Wen$^1$\email, L. Shan$^1$, H.P. Yang$^1$, X. F. Lu$^1$, H. Gao$^1$, Min-Seok Park$^2$, C.U. Jung$^2$, and Sung-Ik Lee$^2$}

\affiliation{$^1$ National Laboratory for Superconductivity,
Institute of Physics, Chinese Academy of Sciences, P.~O.~Box 603,
Beijing 100080, P.~R.~China}

\affiliation{$^2$ National Creative Research Initiative Center for
Superconductivity and Department of Physics, Pohang University of
Science and Technology, Pohang 790-784, Republic of Korea}

\date{\today}

\begin{abstract}
Low temperature specific heat of the electron-doped (n-type)
infinite-layer cuprate $Sr_{0.9}La_{0.1}CuO_{2}$ has been
measured. The quasiparticle density of states (DOS) in the mixed
state is found to be consistent with the feature of the s-wave
pairing symmetry, agreeing very well with the earlier tunnelling
measurement, but being contrary to the d-wave symmetry well
confirmed for the hole-doped (p-type) cuprates. Our results
indicate that the electronic DOS are mainly contributed by the
vortex cores in the present sample being contrast to the p-type
cuprates in which the vortex cores are abnormal and contribute
very limited low energy DOS as evidenced by many means.
\end{abstract}

\pacs{74.20.Rp, 74.25.Bt, 74.25.Fy, 74.72.Dn}

\maketitle The pairing symmetry is one of the essential points in
clarifying the underling physics of high temperature
superconductors (HTS), since it is supposed to be related to the
pairing mechanism in these materials. In the hole doped side, the
pairing symmetry of the cuprate is widely believed to be
$d_{x^2-y^2}$. This has been supported by tremendous
experiments\cite{Tsuei1} both from surface
detection\cite{Tsuei2,ARPES,Hardy,Tunneling,YehNC} and bulk
measurements\cite{NMR,Moler,Revaz,Wright}. For the electron doped
systems, the symmetry of the order parameter remains highly
controversial. Angle-resolved photoemission spectroscopy ( ARPES
)\cite{ShenZX} and phase-sensitive scanning SQUID
measurements\cite{Tsuei3} indicate a d-wave symmetry in
$Nd_{2-x}Ce_xCuO_4$ (NCCO) and $Pr_{2-x}Ce_xCuO_4$ (PCCO). In
addition, Raman scattering shows a nonmonotonic d-wave order
parameter\cite{Blumberg}. Specific heat measurement on $PCCO$
reveal also lines of nodes on the gap function\cite{Greene}.
However, this has been contrasted by tunnelling\cite{YehNC2} and
penetration depth measurement\cite{Alff,Lemberger1}. It has been
recently argued that there may be a crossover from d-wave to
s-wave symmetries by changing the doped electron
concentration\cite{Lemberger2,Biswas}. Recently Chen et
al.\cite{YehNC2} reported evidence of strongly correlated s-wave
pairing in the infinite-layer superconductor
$Sr_{0.9}La_{0.1}CuO_{2}$ by tunnelling spectroscope. This
conclusion can also be corroborated indirectly by the stronger
suppression of $T_c$ by magnetic quantum impurities $Ni$ than
non-magnetic ones $Zn$\cite{SILee}, which is the same as observed
in conventional s-wave superconductors. Since tunnelling technique
relies on the surface situation, a bulk evidence is thus strongly
desirable for the pairing symmetry in this material. It is well
known that the specific heat is one of the important means to
explore the low energy excitations which reflect the bulk
properties\cite{Hussey}. In this Letter we present magnetic field
dependent specific heat of $Sr_{0.9}La_{0.1}CuO_{2}$ in low
temperature region. The magnetic field induced quasiparticle DOS
have been found to be well consistent with a s-wave pairing
symmetry. We further conclude that the vortex cores contribute the
dominant part of DOS in this n-type cuprate. This is in sharp
contrast to the observations in p-type cuprates.

The sample studied in this work is high-density granular material
of $Sr_{0.9}La_{0.1}CuO_{2}$\cite{SILee2}. X-ray diffraction (XRD)
patterns show no any trace of a second phase in the sample. The ac
susceptibility measured with $H_{ac}=1 Oe$ and $f=333Hz$ is shown
in the inset of Fig.1, and one can see that the sample has a sharp
transition at a temperature of $T_c = 43K$. A piece about 23.13 mg
in mass, $2.3 \times 2.0 \times 0.8 mm^3$ in dimensions, was
chosen for the specific heat (SH) measurement. The heat capacity
data presented here were taken with the relaxation method based on
an Oxford cryogenic system Maglab. The heat capacity is determined
by a direct measurement of the thermal time constant,
$\tau=(C+C_{add})/\kappa_w$, here $C$ and $C_{add}$ are the heat
capacity of the sample and addenda (including a small sapphire
substrate, small printed film heater, tiny Cernox temperature
sensor, $\phi$25 $\mu m$ gold wire leads, Wakefield thermal
conducting grease ($100\mu g$)) respectively, where $\kappa_w$ is
the thermal conductance between the chip and a thermal link. The
value $C_{add}$ has been measured and subtracted from the total
heat capacity, thus $C$ value reported here is only the heat
capacity of the sample. We have also checked the field dependence
of $C_{add}$ and found that the change (if any) of $C_{add}$ under
12 T is in the same order of the noise background (20 nJ/K at 5 K
and 40 nJ/K at 20 K). The influence of the magnetic field (12 T)
on the readout of the thermometer is below 0.02 K and can be
neglected. During the measurement, the sample was cooled to the
lowest temperature under a magnetic field ($H||c$) (field-cooling)
followed by data acquisition in the warming up process.

\begin{figure}
\includegraphics[width=8cm]{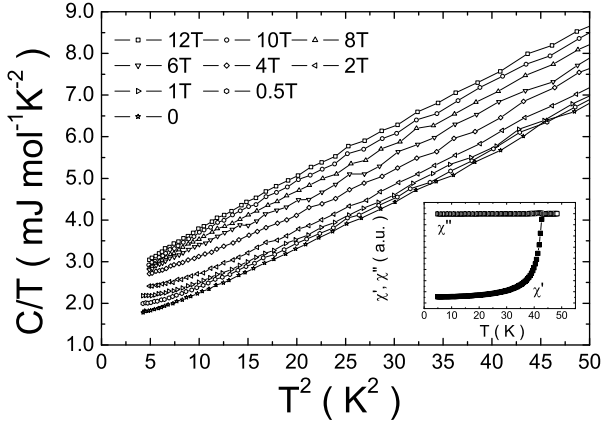}
\caption{Specific heat coefficient C/T vs. $T^2$ at magnetic
fields ranging from 0 to 12 T for the sample
$Sr_{0.9}La_{0.1}CuO_{2}$. The inset shows the ac susceptibility
of the sample.} \label{fig1}
\end{figure}

Fig.1 shows the SH coefficient $C/T$ vs. $T^2$ at magnetic fields
ranging from 0 to 12 T. The separation between each field can be
well determined. In low temperature region the curves are upturned
below 4 T, and gradually a slight but broad hump appears for
higher magnetic fields, these are due to the Schottky anomaly of
free spins which will be discussed later. Beside the Schottky
anomaly, the curve at zero field extrapolates to a finite value
$\gamma_0 \approx 1.2 mJ/mol K^2$ at 0 K, which is about 3 to 5
times smaller than that observed in p-type cuprate
superconductors. If the sample has the d-wave symmetry, this can
be interpreted as potential scattering near the node of
$d_{x^2-y^2}$ gap function due to small amount
impurities\cite{Moler}. However, if the sample has a typical
s-wave symmetry, this term should be very weak if the scatterers
are non-magnetic. The small but finite value of $\gamma_0$
observed here may be still explained as due to impurity scattering
with strong correlation effect in the present sample although it
is s-wave\cite{YehNC2}. In addition, for the temperature above 8
K, the SH deviates from a $T^3$ law due to higher order terms in
the phonon SH. To avoid extra parameters needed to describe the
phonon SH above 8K, our analysis is restricted to T$\leq$7K.

It is known that different gap symmetries give rise to different
quasiparticle DOS $N_F$ near the Fermi level\cite{Hussey}.
Conventional low-$T_c$ superconductors show an s-wave gap symmetry
in which the electronic SH has an exponential temperature
dependence, $C_{el}\propto Te^{-\Delta/k_BT}$, where $\Delta$ is
the energy gap. For present sample, $\Delta$ is about 13
meV\cite{YehNC2}, thus $C_{el}$ can be negligible at temperatures
considered here ($T\leq 7 K$). For a clean d-wave superconductor
with nodes, since $N_F\propto E$ near the Fermi level, this leads
to the relation $C_{el}\propto T^{2}$\cite{Volovik}. Therefore we
fit the zero field SH data with Eq.(1) and Eq.(2) according to
d-wave and s-wave ( when $k_BT<<\Delta$ ) respectively,
\begin{equation}
C(T,0)=\Theta/T^{2}+\gamma_{0}T+\alpha{T^2}+\beta{T^3}
\end{equation}
\begin{equation}
C(T,0)=\Theta/T^{2}+\gamma_{0}T+\beta{T^3}
\end{equation}
where $\Theta/T^2$ is the zero field Schottky anomaly due to the
effective internal field, $\beta{T^3}$ is the phonon term,
$\gamma_{0}T$ is the zero field linear term as discussed above,
and the $\alpha{T^2}$ term is due to the excitations with lines of
nodes in d-wave case. The fit results of the zero field data to
Eq.(1), with the phonon coefficient $\beta$ to be free and to be
fixed as the average of the fit results of the high fields, $0.118
mJ/mol K^4$, are shown in Table I.

\begin{center}
Table I. Zero-field specific heat fit to Eq.(1).

(Units are mJ,mol, and K.)
\begin{tabular}
{c cc cc cc cc} \hline\hline \multicolumn{1}{c}{H}&
\multicolumn{2}{c}{$\Theta$}& \multicolumn{2}{c}{$\gamma_0$}&
\multicolumn{2}{c}{$\alpha$}& \multicolumn{2}{c}{$\beta$}
\\ \hline $0.0$ & $1.68\pm0.28$& &\multicolumn{2}{c}{$1.29\pm0.07$}&\multicolumn{2}{c}{$-0.10\pm0.03$}&\multicolumn{2}{c}{$0.123\pm0.003$}
\\\hline $0.0$
 & \multicolumn{2}{c}{$2.0\pm0.1$} & \multicolumn{2}{c}{$1.18\pm0.02$}& \multicolumn{2}{c}{$-0.054\pm0.003$}& \multicolumn{2}{c}{$0.118$}
\\ \hline
\end{tabular}
\end{center}

From table I, one can see that the zero field data can not produce
reasonable $\alpha$ both with free $\beta$ and with the fixed
$\beta$. This indicates that the zero field DOS of the sample does
not have the term $\alpha{T^2}$. So we use Eq.(2), the case of
s-wave when $k_BT<<\Delta$, to fit the zero field data and deduce
the reasonable value of $\beta$. The absence of the $\alpha T^2$
may indicate preliminarily that the pairing symmetry of present
sample is not clean d-wave like.

Then we take a general fit to the SH data at different fields. No
matter the sample has a s-wave or d-wave symmetry, at a fixed
magnetic field, in low temperature region the fit formulae can be
written as\cite{Moler},

\begin{equation}
C(T,H)=C_{Sch}(T,H)+\gamma(H)T+\beta T^3
\end{equation}
\begin{equation}
C_{Sch}(T,H)=n(\frac{g\mu_{B}H}{k_{B}T})^2\frac{e^{g\mu_{B}{H}/k_{B}{T}}}{(1+e^{g\mu_{B}H/k_{B}{T}})^{2}}
\end{equation}

where $C_{Sch}(T,H)$ is the Schottky anomaly under a magnetic
field; n is related to the concentration of spin-1/2 particles;
and $\gamma(H)T$ is the sum of the zero field linear term and the
magnetic field induced linear contribution, $\Delta
\gamma$=$\gamma(H)-\gamma_0$ $\propto H$ for s-wave and $\propto
\sqrt{H}$ for d-wave. All data are fit with a Land$\acute{e}$ $g$
factor of $g=2.0$ and the fit results are shown in Table II.

\begin{center}
Table II. Fit of specific heat at fixed magnetic fields to Eq.(3)

from 2-7K. (Units are mJ, K, and T)
\begin{tabular}
{c cc cc cc} \hline\hline \multicolumn{1}{c}{H}&
\multicolumn{2}{c}{$\gamma$}& \multicolumn{2}{c}{$\beta$}&
\multicolumn{2}{c}{n}
\\ \hline $0.0$ & $1.04\pm0.01$& &\multicolumn{2}{c}{$0.1126\pm0.0004$}&\multicolumn{2}{c}{$$}
\\ \hline $0.5$ & $1.13\pm0.02$& &\multicolumn{2}{c}{$0.1136\pm0.0006$}&\multicolumn{2}{c}{$33.3\pm2.1$}
\\ \hline $1.0$ & $1.21\pm0.02$& &\multicolumn{2}{c}{$0.1139\pm0.0006$}&\multicolumn{2}{c}{$11.3\pm0.6$}
\\ \hline $2.0$ & $1.34\pm0.02$& &\multicolumn{2}{c}{$0.1158\pm0.0006$}&\multicolumn{2}{c}{$4.5\pm0.2$}
\\ \hline $4.0$ & $1.59\pm0.04$& &\multicolumn{2}{c}{$0.1186\pm0.0010$}&\multicolumn{2}{c}{$2.8\pm0.2$}
\\ \hline $6.0$ & $1.77\pm0.11$& &\multicolumn{2}{c}{$0.1195\pm0.0022$}&\multicolumn{2}{c}{$3.1\pm0.7$}
\\ \hline $8.0$ & $2.09\pm0.05$& &\multicolumn{2}{c}{$0.1199\pm0.0007$}&\multicolumn{2}{c}{$2.1\pm0.5$}
\\ \hline $10.0$ & $2.25\pm0.02$& &\multicolumn{2}{c}{$0.1214\pm0.0004$}&\multicolumn{2}{c}{$2.7\pm0.3$}
\\ \hline $12.0$ & $2.41\pm0.01$& &\multicolumn{2}{c}{$0.1215\pm0.0004$}&\multicolumn{2}{c}{$3.1\pm0.3$}
\\ \hline
\end{tabular}
\end{center}

\begin{figure}
\includegraphics[width=8cm]{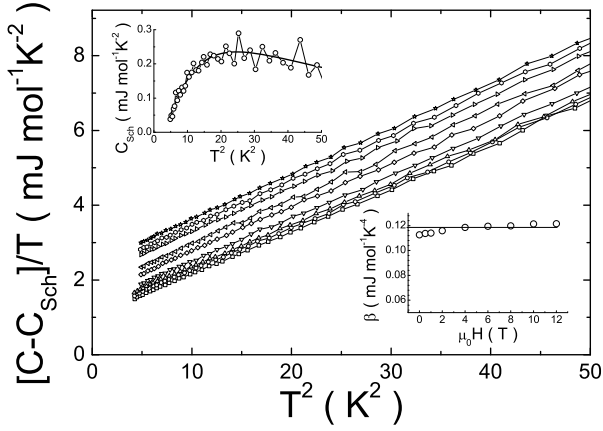}
\caption{Temperature dependence of the specific heat at various
fields with the Schottky terms subtracted. The bottom-right inset
shows the phonon coefficient $\beta$ vs H with the horizontal line
marking the value $\beta$=0.118. The upper-left inset shows a
typical example for the result of the Schottky term (12 T)
$C_{Sch}=C-\gamma(H)T-\beta T^3$, and the solid line represents a
fit to Eq.(4). } \label{fig2}
\end{figure}

The fit results of the phonon term $\beta$ and the Schottky term
at 12 T are shown as insets of Fig.2, where
$C_{Sch}=C-\gamma(H)T-\beta T^3$. The large value $n$ in low field
region may be induced by the residual effect of the internal
crystal field. In order to remove the influence of the Schottky
anomaly, we subtract the raw data with $C_{Sch}$ and the results
are shown in the main panel of Fig.2.

In the mixed state, there are two types of quasiparticle
excitations in the bulk of a superconductor: bound states inside
the vortex cores, and extended states outside the vortex cores (
as predicted for d-wave \cite{Volovik} ). In conventional s-wave
superconductors, the inner core bound states dominate the
quasiparticle excitations; therefore, the electronic SH is
proportional to the number of vortices. The number of vortices
increases linearly with field, thus the magnetic field induced
electronic SH is proportional to H\cite{Hussey}, i.e.,
$C_{core}\approx \gamma_nTH/H_{c2}(0)$. If we divide each side by
$T^3$, one obtains

\begin{equation}
\frac{C_{core}}{T^3}=\frac{\gamma_n}{H_{c2}(0)}(\frac{T}{\sqrt{H}})^{-2}
\end{equation}

For a gap with lines of nodes (e.g., d-wave symmetry), the
extended quasiparticles dominate the excitation spectrum in the
clean limit. It has been shown that the electronic SH has a
$\sqrt{H}$ dependence in the clean limit at $T=0$\cite{Volovik}
and the data should obey Simon-Lee\cite{SimonLee} scaling law

\begin{equation}
\frac{C_{vol}}{T^{2}}=f(\frac{T}{\sqrt{H}})
\end{equation}

Since the phonon SH is field independent and the Schottky
contribution has been removed from the raw data, we can obtain the
field dependent part of the electronic SH through subtracting the
zero field SH from the one measured at other fields. Therefore for
the s-wave symmetry,
$C_{cal-s}=[(C(H)-C_{Sch}(H))-(C(H=0)-C_{Sch}(H=0))]/T^3 \propto
H/T^2$, thus should scale with $T/\sqrt{H}$; for the d-wave
symmetry,
$C_{cal-d}=[(C(H)-C_{Sch}(H))-(C(H=0)-C_{Sch}(H=0))]/T^{2} =
C_{vol}/T^2-\alpha$, should also scale with $T/\sqrt{H}$.

\begin{figure}
\includegraphics[width=8cm]{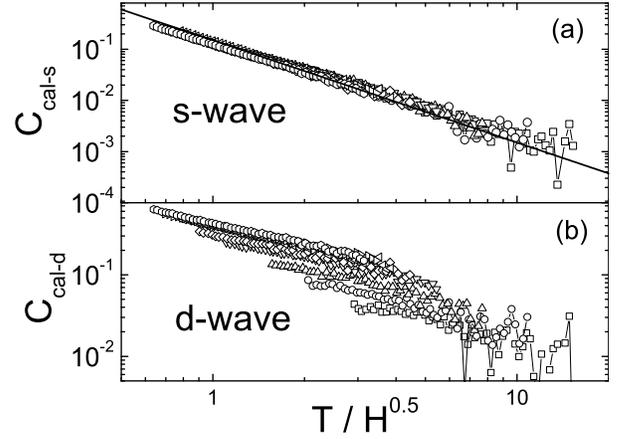}
\caption{a. Scaling of the data (symbols)
$C_{cal-s}=[(C(H)-C_{Sch}(H))-(C(H=0)-C_{Sch}(H=0))]/T^{3}$ vs.
T$/\sqrt{H}$, the solid line represents the theoretical expression
Eq.(5); b. Scaling of the data (symbols)
$C_{cal-d}=[(C(H)-C_{Sch}(H))-(C(H=0)-C_{Sch}(H=0))]/T^{2}$ vs.
T$/\sqrt{H}$. No good scaling can be found for the d-wave case.}
\label{fig3}
\end{figure}

The scaling result of the magnetic field induced DOS with the
s-wave condition (Eq.(5)) is presented in Fig.3a. The scaling
quality is quite good. The solid line is a theoretical curve
$C_{cal-s}=0.15H/T^2$, thus $\gamma_n/H_{c2}(0)\approx 0.15 mJ/mol
K^2T$ according to Eq.(5). Taking $H_{c2}(0)$=50T, $\gamma_n=7.5
mJ/molK^2$, which is quite close to the value of optimally doped
$LSCO$\cite{Nohara}. Fig.3b shows the scaling by following the
d-wave condition (Eq.(6)). It is obvious that the result of the
scaling for the s-wave condition is much better than the clean
d-wave condition. Therefore it is tempting to conclude that the
pairing symmetry of the sample is not the clean d-wave but very
likely s-wave. The field induced quasiparticle DOS are contributed
mainly by the vortex cores.

Further more, the zero temperature electronic DOS $\gamma(T=0)(H)$
was obtained and shown in Fig.4. The circles represent raw data,
and the solid line is the fit using the formula
$\gamma=\gamma(0)+AH^B$. It is known that the expected value for
$B$ is 0.5 for d-wave and 1 for s-wave. The obtained $B$ is 0.85,
being close to the linear condition B=1. It is important to note
that all SH measurements on p-type cuprates (mostly near optimal
doped point) yield a value $B\approx 0.5$ giving the evidence for
d-wave. Distinction from the d-wave symmetry is apparent in
present sample although $B=0.85$ instead of 1 is found here.
Actually the curve between 1 T and 10 T is close to be linear. It
has been pointed out that\cite{Ichioka} when the field is
relatively high, the vortex lattice effect should be considered
and usually the linear H dependence of the quasiparticle DOS
cannot be seen for s-wave. In Fig.4 the dotted line is a fit to
the clean d-wave superconductor with $B= 1/2$. It is clear that
the clean limit d-wave cannot describe the data at all. We have
also used the dirty-limit relation to fit the zero temperature
DOS. At the unitary limit and T = 0, K$\ddot{u}$bert and
Hirschfeld\cite{Kubert} predicted that the field induced DOS for
d-wave is $\delta\gamma/\gamma_0=P_1(H/P_2)log(P_2/H)$, where
$P_1=0.322(\Delta_0/\Gamma)^{1/2}$, $\Delta_0 $ the gap maximum,
$\Gamma$ the impurity scattering rate, and $P_2=\pi H_{c2}/2a^2$,
$a\approx 1$. Worthy of noting is that this relation is too
flexible which can apparently fit to data with strong diversity.
The fit to our data yields $\pi H_{c2}/2a^2$=3604. Taking a=1, one
has $H_{c2}=2294 T$ which is far beyond the reasonable value. So
the d-wave in the dirty limit cannot be used to interpret our data
either. This is in sharp contrast with what appears for the p-type
cuprates. Recently SH measurements on an overdoped LSCO single
crystal (x=0.22) revealed also a d-wave pairing symmetry and very
limited contribution from the vortex cores to the electronic DOS
in the mixed state\cite{LiuZY}. All these indicate that the
features of vortex cores in present sample are very different from
that in p-type ones in which a gapped electronic state may appear
within the vortex cores when the superconductivity is
suppressed\cite{STMCore,NMRCore}.

In summary, specific heat measurements reveal that the electronic
DOS in the n-type infinite-layer superconductor
$Sr_{0.9}La_{0.1}CuO_2$ are mainly contributed by the vortex cores
showing a bulk evidence for the s-wave pairing symmetry. This
manifests that the d-wave pairing symmetry may not be universal in
cuprate superconductors, rather it depends on the specific
structure and competing ground states.

\begin{figure}
\includegraphics[width=8cm]{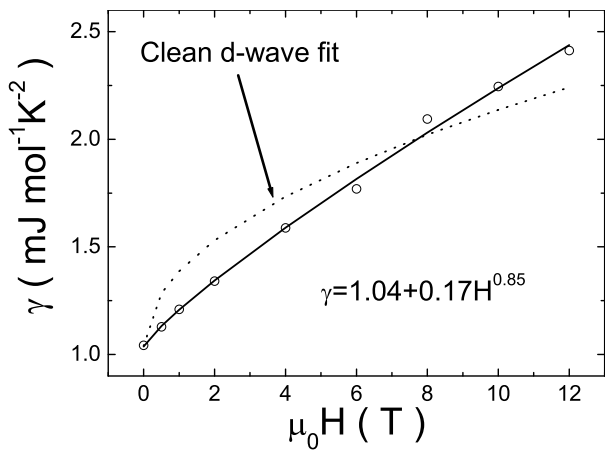}
\caption{Field dependence of the coefficient of the linear-T term
$\gamma$ at zero temperature. The solid line represents the fit to
$\gamma=\gamma(0)+AH^{B}$, and the dotted line for the clean
d-wave case. } \label{fig4}
\end{figure}

\section{Acknowledgments}

This work is supported by the National Science Foundation of China
(NSFC), the Ministry of Science and Technology of China, the
Knowledge Innovation Project of Chinese Academy of Sciences. The
work in Pohan University is supported by the Ministry of Science
and Technology of Korea through the Creative Initiative Program.

% \appendix

% \section{}
% \label{}

Correspondence should be addressed to hhwen@aphy.iphy.ac.cn


\begin{thebibliography}{00}

\bibitem{Tsuei1} C. C. Tsuei, and J. R. Kirtley, Rev. Mod. Phys. {\bf72}, 969
(2000), and references therein.
\bibitem{Tsuei2} C. C. Tsuei, et al. Nature {\bf387}, 481
(1997).
\bibitem{ARPES} Z.-X. Shen, et al. Park, Phys. Rev. Lett. {\bf70},
1553 (1993); D. J. Scalapino, Phys. Rep. {\bf250}, 330 (1995).
\bibitem{Hardy} W. N. Hardy, et al. Phys. Rev. Lett. {\bf70}, 3999 (1993).
\bibitem{Tunneling} A. G. Sun, et al. Phys. Rev. Lett. {\bf72}, 2267 (1994).
\bibitem{YehNC} N.-C. Yeh, et al. Phys. Rev. Lett. {\bf87}, 87003 (2001).
\bibitem{NMR} N. Bulut and D. J.
Scalapino, Phys. Rev. Lett. {\bf68}, 706 (1992). G.-q. Zheng, et
al. Phys. Rev. Lett. {\bf88}, 77003 (2002).
\bibitem{Moler} K. A. Moler, et al. Phys. Rev. Lett. \textbf{73}, 2744 (1994). K. A. Moler, et al. Phys. Rev.
B {\bf55}, 12753 (1997).
\bibitem{Revaz} B. Revaz, et al. Phys. Rev. Lett. {\bf80}, 3364 (1998).
\bibitem{Wright} D. A. Wright, et al. Phys. Rev. Lett. {\bf82}, 1550 (1999).
\bibitem{ShenZX} N. P. Armitage, et al., Phys. Rev. Lett. \textbf{86},
1126 (2001).
\bibitem{Tsuei3} C. C. Tsuei and J. R. Kirtley, Phys. Rev. Lett.
\textbf{85}, 182 (2000).
\bibitem{Blumberg} G. Blumberg, et al., Phys. Rev. Lett. \textbf{88}, 107002 (2002).
\bibitem{Greene}H. Balci, et al., Phys. Rev. B \textbf{66}, 174510
(2002).
\bibitem{YehNC2} C. T. Chen, et al., Phys. Rev. Lett. \textbf{88}, 227002
(2002).
\bibitem{Alff} L. Alff, et al., Phys. Rev. Lett. \textbf{83}, 2644
(1999).
\bibitem{Lemberger1} J. A. Skinta, T. R. Lemberger, Phys. Rev. Lett.
\textbf{88}, 207003 (2002).
\bibitem{Lemberger2} J. A. Skinta, M. S. Kim, and T. R. Lemberger, Phys. Rev. Lett. \textbf{88}, 207005
(2002).
\bibitem{Biswas} A. Biswas, et al., Phys. Rev. Lett. \textbf{88},
207004 (2002).
\bibitem{SILee}C. U. Jung, et al. Phys. Rev. B \textbf{65}, 172501 (2002).
\bibitem{Hussey}For a recent review, please see N. E. Hussey, Adv.
in Phys.\textbf{51},1685 (2002).
\bibitem{SILee2}C. U. Jung, et al. Current Appl. Phys. \textbf{1}, 157 (2001).
\bibitem{Volovik} G.E. Volovik, JETP Lett. {\bf 58}, 469 (1993);
ibid {\bf65}, 491 (1997).
\bibitem{SimonLee}S. H. Simon, P. A. Lee, Phys. Rev. Lett. {\bf78},
1548 (1997).
\bibitem{Nohara}M. Nohara, et al., J. Phys. Soc. Jpn.\textbf{69},
1602(2000).
\bibitem{Kubert} C. K$\ddot{u}$bert, P. J. Hirschfeld, Solid State Commun {\bf105}, 459
(1998).
\bibitem{Ichioka} M. Ichioka, A. Hasegawa, and K.
Machida, Phys. Rev. B, {\bf59}, 184 (1999).
\bibitem{LiuZY}Z. Y. Liu, et al. Condmat: 0301366.
\bibitem{STMCore} I. Maggio-Aprile, et al., Phys. Rev. Lett.{\bf 75}, 2754 (1995); Ch. Renner, et al., Phys. Rev. Lett. {\bf80}, 3606 (1998); S. H. Pan, et al., Phys. Rev. Lett. {\bf85}, 1536 (2000); B. W. Hoogenboom, et al., Phys. Rev. Lett. {\bf87}, 267001 (2001).
\bibitem{NMRCore} V. F. Mitrovic, et al., Nature {\bf 413}, 501 (2001).

%%%%%%%%%%%%%%%%%%%%%%%%%%%%%%%%%%%%%%%%%%%%%%%%%%%%%%%%%%%%%%%%%%%%








% notes:

% \bibitem{label2}
% If there is a note, it should come last:
% \bibitem{label3} \note
% \end{subbibitems}


\end{thebibliography}
\end{document}